\documentclass[useAMS,usenatbib,letters]{mn2e}
\usepackage{aas_macros,graphicx,times,multirow}
\usepackage{amssymb}
\title[The quiescent emission of \xte]
{Emission geometry, radiation pattern, and magnetic topology of the
magnetar \xte\ in its quiescent state}
\author[F. Bernardini et al.]                                                    
{F.~Bernardini,$^{1,2,3}$\thanks{E-mail: bernardini@mporzio.astro.it} R.~Perna$^{4}$, E. V. Gotthelf$^{5}$, G. L. Israel$^{2}$, 
N. Rea$^{6}$, and L. Stella$^{2}$\\
$^1$ Universit\`a degli Studi di Roma ``Tor Vergata" Via Orazio Raimondo 18, I-00173 Roma, Italy\\
$^2$ INAF $-$ Osservatorio Astronomico di Roma, Via Frascati 33, I-00040 Monteporzio Catone (Roma), Italy\\
$^3$ INAF $-$ Osservatorio Astronomico di Capodimonte, Salita Moiairiello 16, I-80131 Napoli, Italy\\
$^4$ JILA, Univ. of Colorado, Boulder, CO 80309$-$0440, USA\\
$^5$ Columbia Astrophysics Laboratory, Columbia University, 550 West 120th Street, New York, NY 10027-6601\\
$^6$ Institut de Ciencies de l'Espai (ICE, CSIC-IEEC), 08193, Barcelona, Spain}
\date{}
\pagerange{\pageref{firstpage}--\pageref{lastpage}}
\def\xte{XTE\,\,J1810$-$197}

\def\e15{1E\,1547.0-5408}

\def\sg16{SGR\,1627-41}
\def\cxou{CXOU J164710.2$-$455216}

\def\XMM{{\em XMM-Newton}}

\def\ergscm{\rm erg\,cm^{-2}\,s^{-1}}

\newcommand\beq{\begin{equation}}
\newcommand\eeq{\end{equation}}
\def\rmmat#1{{\hbox{\rm #1}}}

\def\taxp{\hbox{XTE~J1810$-$197}}

\begin{document}
\label{firstpage}
\maketitle
\begin{abstract}
The return to the quiescent state of the Anomalous X-ray pulsar \xte\
following its 2003 outburst represents a unique opportunity to probe
the surface emission properties of a magnetar.  The
quiescent emission of \xte\ is composed of two thermal components, one arising
from the whole star surface, and the other from a small warm spot on
it. By modeling the magnitude and shape of the pulse profile in narrow
spectral bands, we have been able to constrain the physical
characteristics and geometrical parameters of the system: the two
angles that the line of sight and the spin axis make with respect to
the warm spot axis ($\psi$ and $\xi$ respectively), the angular size
of the spot, and the overall surface temperature distribution.  Our
modeling accounts for the general relativistic effects of
gravitational redshift and light bending near the stellar surface, and
allows for local anisotropic emission.  We found that the surface
temperature distribution on the neutron star is consistent with the
expectations of a dipole magnetic field configuration; the local
radiation requires a pencil-beamed emission pattern, suggesting the
presence of a magnetized atmosphere.  For a typical value of the
radius, R=13~km, the viewing parameters (symmetric for an interchange
between $\psi$ and $\xi$), range from $\psi=\xi=38^{\circ}$ to
($\psi,\xi$)=($52^{\circ}$,$29^{\circ}$). These angles are consistent
with those obtained by modeling the AXP in outburst, with uncertainty
contours reduced by a factor of 2.5.
\end{abstract}
\begin{keywords}
pulsars: general -- stars: neutron -- X-rays: individual: \xte.
\end{keywords}

\section{Introduction}

\xte\ is an isolated neutron star (NS) belonging to the class of the
Anomalous X-ray Pulsars (AXPs); these objects, together with the Soft
Gamma-ray repeaters (SGRs), are believed to be magnetars: isolated
neutron stars whose thermal emission and occasional outbursts
are powered by their extremely strong magnetic fields (Duncan \&
Thompson 1992; Thompson \& Duncan 1995).  Originally one of the
thousands of the faint X-ray sources cataloged by ROSAT, the variable
nature of \xte\, was revealed by the outburst of 2003, with its sudden
increase in X-ray luminosity by a factor $\sim 100$, decaying on a
time-scale of years.  The source rotation period and its derivative
were consequently measured and found to be $P=5.54\,\rm s$ and
$\dot{P}=1.1-2.1\times10^{-11}\,\rm s/s$. These timing properties
imply a magnetic field $B_{\rm dip}\sim 3\times10^{14}\,\rm
G$, confirming the magnetar classification of the source (Ibrahim et
al. 2004, Gotthelf at. al. 2004). The source was monitored repeatedly
for more than 7 years with several X-ray observatories, up to the
return to quiescence (Gotthelf \& Halpern 2005, 2007; Halpern \&
Gotthelf 2005;  Bernardini et al. 2009;  Albano et
al 2010).  These studies presented a unique opportunity to probe the
emission mechanisms of a strongly magnetized NS by taking advantage
of the flux evolution during its decay. While analysis of
phase-averaged spectra alone cannot uniquely distinguish among
competing emission models, the addition of the steady change of the
spectrum and pulse profile over time greatly increases the diagnostic
power.

Perna \& Gotthelf (2008) developed a detailed emission model for the
energy-dependent pulse profile of \xte\, following its outburst. This
model, which was tailored to the specific surface emission
distribution in the post-outburst phase, can take on any viewing
geometry, includes the general relativistic effects of light
deflection and gravitational redshift, and allows for local
anisotropic emission. The application of this model to the first 4
sets of \XMM\ data acquired during the temporal evolution of the flux
from \xte\ following the outburst (September 2003 - September 2004),
provided a constraint on the underlying emission geometry and
radiation properties of this transient magnetar in its post-outburst
phase.

\begin{figure}
\centering
\includegraphics[scale=0.45,angle=270]{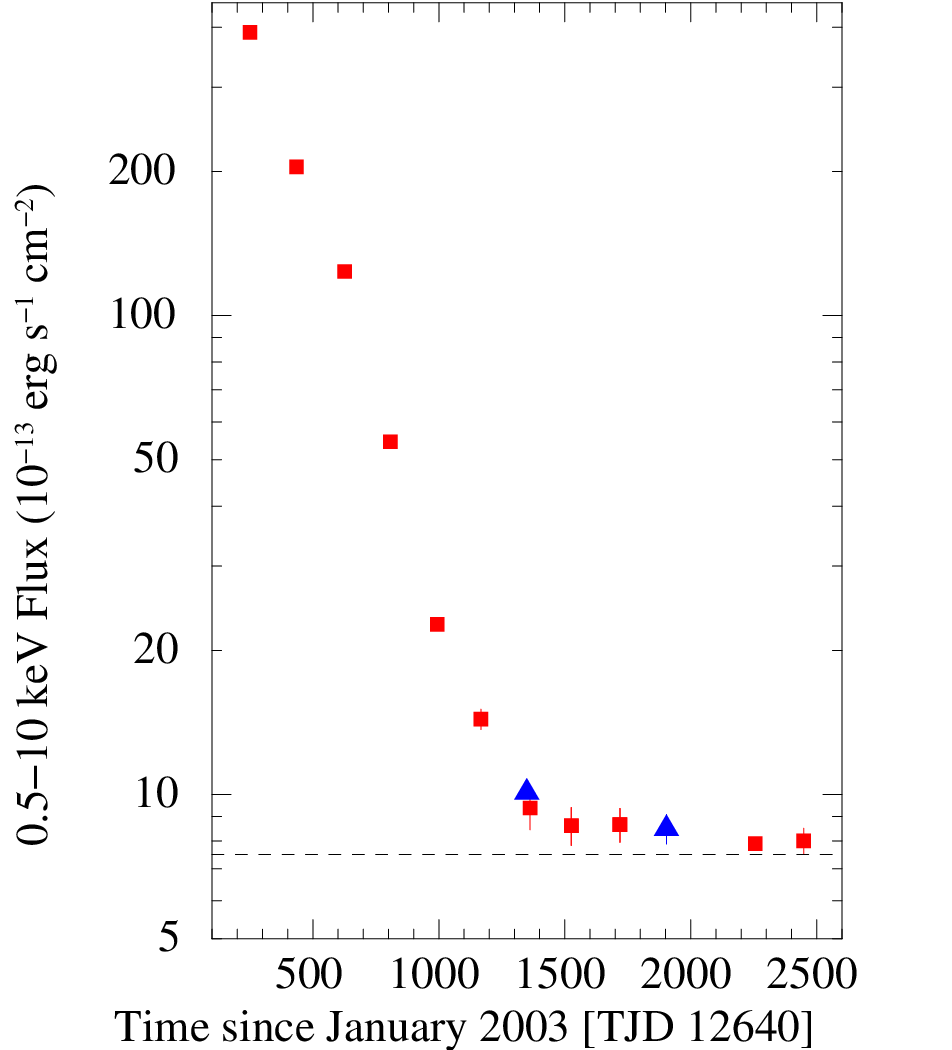}
\caption{{\em XMM} (red) and {\em Chandra} (blue) 0.5--10 keV 
flux measurements of \xte\ following its outburst.  It is evident
from the data that \xte\ has now reached quiescence. The dashed line
represents the X-ray flux level ($\sim7.5\times10^{-13}\,\ergscm$) as
recorded by $ROSAT$, Einstein, and $ASCA$ before the outburst
onset (Ibrahim et al. 2004, Gotthelf et al. 2004). 
The analysis presented in the paper is performed over three
close \XMM\ pointings represented by the last data point.}
\label{fig:fluxevol}
\end{figure}

In this paper we present the results of the modeling of the spectral
and timing data of \xte\ obtained with 3 combined \XMM\ pointings
(September 2009) upon the return of the source to quiescence (see
Figure \ref{fig:fluxevol}, last point). Indeed, our goal here is that
of studying the properties of the quiescent emission of this magnetar,
which carries information on the surface temperature distribution of
the star, and hence on its magnetic topology. We model these data
using a modified version of the emission model by Perna \& Gotthelf
(2008), updated to include: {\em (a)} the presence of quiescent
emission from the full surface of the star, {\em (b)} the changed and
reduced emission from the region heated by the outburst.

The new data and its spectral and timing analysis is reported in
\S2. In \S3, we discuss the properties of the quiescent
emission. The theoretical emission model is described in detail in
\S4, and the results of its application to the data are given in \S5,
following with a discussion in \S6.

\section{Observations}
\label{sec:obs}

The quiescent state of the source was observed by \XMM\ during three
consecutive close pointings (18 days total time span) in 2009, on
September 5, 7, 23 for $19, 18, 12$~ks, respectively. All the
observations were performed with the PN instrument (Str\"uder et
al. 2001) in large window mode, with the medium filter applied, and
the MOS 1 and MOS 2 instruments (Turner et al. 2001) in small window
mode, with the use of the medium and thin filter, respectively. The PN
resolution time with this configuration is $47.6$~ms while the MOS1/2
resolution time is 0.3~s.  Data were processed with SAS version
10.0.0, using the updated calibration files (CCF) available in August
2010. Standard data screening criteria were applied in the extraction
of scientific products. Time window criteria were used for removing
time intervals contaminated by solar flares. A total of 48~ks 
of good exposure time was obtained. Photon arrival times were
converted into barycentric dynamical times (TBD) using the SAS tool
{\tt barycen} and the milliarcsec radio position of Helfand et
al. (2007): RA=18$^h$ 09$^m$ 51$\fs$0870, Dec=-19$^{\rm o}$ 43$^{'}$
51$^{''}.$931 (J2000).

Source photons were extracted from a centered circular region of
radius 55\arcsec\ containing 90\% of the source counts. For both the
timing and spectral analysis we extracted the background from the same PN
or MOS CCD where the source lies using a circular region of the same
size as that of the source. In the following analysis, we
combine the three PN spectra collected by \XMM\ over an interval of
$18$ days, after verifying that the individual spectra were consistent
with each other within the uncertainties. The spectra were binned to
have at least 30 counts per fitting channel to insure adequate fit
statistics.

For each observation and instrument we extracted and summed data into
four energy bands (following Perna \& Gotthelf, 2008): 0.5$-$1,
1$-$1.5, 1.5$-$2 and 2$-$3 keV. Lightcurves were generated at the
minimum allowed bin time of 0.3~s. We phase connected the three
multi-instrument lightcurves using the phase-fitting technique
outlined in Dall'Osso et al. (2003).  The timing solution, referred to
epoch MJD 54079, includes only one term, the rotation period, which
was found to be $P=5.5406556\pm2\times10^{-7}$~s.  The brief span of
the observations did not allow for a spin-down measurement, but only
for a 3$\sigma$ upper limit of $\dot{P}\leq 9\times 10^{-12}$~s/s
(consistent with the value $\dot{P}=0.8-1.0\times 10^{-11}$~s/s
reported by Camilo et al. 2007 for the 2006-2007 observation period).
The pulse profile at different energy bands was determined by means of
the $P$ value reported above. The profile was found to be nearly
sinusoidal and energy independent in analogy with earlier epochs
(Halpern and Gotthelf 2005; Bernardini et al. 2009).

\section{Properties of the quiescent emission}
\label{sec:propquie}

\begin{figure*}
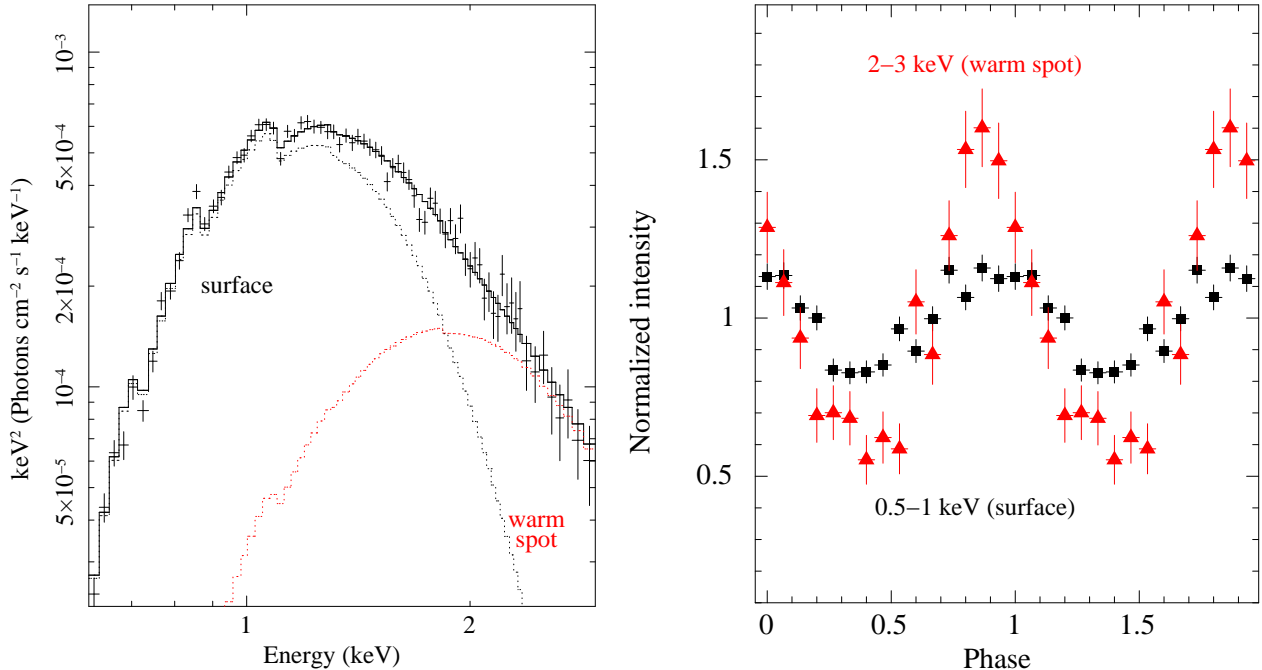

\begin{center}
\begin{tabular}{cc}
\includegraphics[width=0.50\textwidth, angle=-90]{figsurfwarm.ps} & 
\includegraphics[width=0.50\textwidth, angle=-90]{profpuls05123.ps}\\
\end{tabular}
\caption{ {\it Left panel}: The quiescent spectrum of \xte\ modeled
  with a 2BB model with the softer component associated with the 
entire surface and the hotter component, with a higher level of PF, associated 
with a localized warm spot. {\it Right panel}: Pulse profile
  of the surface component (black square) compared to the warm spot
  pulse profile. Both profiles show a peak at the same phase interval
  ($\phi\sim0.85$).}
\label{fig:2bb_prop}
\end{center}
\end{figure*}

Halpern \& Gotthelf (2005) and Gotthelf \& Halpern (2005) showed that
the post-outburst spectrum was composed of a multi-blackbody (BB) made
up of two thermal components, which they interpreted as a warm ring
surrounding a hot spot.  Bernardini et al. (2009) further showed
evidence for a third, cooler thermal component consistent with
emission from the whole surface of the star.  The recorded outburst
flux was a factor $\sim100$ times higher than in the quiescent state.
They also found that, while the intensity of the emission from the
whole NS surface was constant during the outburst (and equal to the
one recorded in quiescence), the flux from the two hotter regions
decreased exponentially with time (on a timescale $\tau\sim1$ yr).
Spectral analysis showed that the warm and the hot regions were
shrinking with time, and the total luminosity of the star was
consequently declining towards the quiescent level. When the source
eventually returned to quiescence, its spectrum was fully consistent
with that recorded by ROSAT before the outburst. More in detail, the
quiescent flux was found to be composed of a cool BB component
consistent with emission from the whole NS surface (for the best
estimated distance value of $\sim3.3$ kpc; Camilo et al. 2006, Minter
et al. 2007, Durant \& van Kerkwijk 2006), and a warmer BB emission
coming from a small, residual spot. The quiescent spectrum of the
source is therefore composed by two BB only (see Figure
\ref{fig:2bb_prop} left panel).  It should however be noted that the
quiescent spectrum also displays a statistically significant
($\sim5\sigma$) absorption feature around 1.1 keV. This feature,
discussed in Bernardini et al. (2009), is of unknown origin. It could
be a proton cyclotron line if the magnetic field is $B=2.2\times
10^{14}$~G (an electron cyclotron line would imply a $B$ field about
2000 times weaker, out of the magnetar range). In our spectral fits,
the feature is modeled with an edge.  We have verified that, within
uncertainties, the absorption depth ($\tau_{c}$) at the energy
threshold ($E_{c}$) is independent of phase, and of magnitude
$<\tau_{c}>=0.32\pm0.02$.  Hence the presence of this feature is not
expected to affect the results from the timing analysis and is not
included therein. Such a feature is however included in all spectral
fits in the present work.

Since \xte\ has now returned to quiescence, the temperature
distribution on the star surface reflects the overall magnetic field
distribution, as the conductivity is enhanced along magnetic field
lines. For most AXPs, the high level of pulsed fraction cannot be
produced by a temperature distribution following a dipole magnetic
field (De Deo, Psaltis \& Narayan 2000). In the case of \xte, on the
other hand, the low level of pulsed fraction of the soft X-ray bands could
be the result of such a field configuration, which we hence adopted as
our starting point.  A determination of the \xte\ surface temperature
distribution would allow to unveil its magnetic field configuration.

The spectral component corresponding to the cold NS
surface easily dominates the emission below 1 keV, while the emission
corresponding to the warm spot dominates over 2 keV (see figure
\ref{fig:2bb_prop} left panel). 
Our first goal was the localization on the star surface of the warm spot
with respect to the maximum intensity of the cooler surface emission
(corresponding to the magnetic axis).  Note that, albeit the NS
surface is expected to have a temperature gradient, however, due to
the limited S/N ratio of the spectral data, it can only be identified
as a single BB in the spectral analysis (i.e. the colder one).  
The pulse profiles of the two BB components were 
consequently generated separately by taking into account
only photons emitted in the 0.5$-$1 keV and 2$-$3 keV energy bands
respectively.  The result of this test shows (see figure
\ref{fig:2bb_prop} right panel) that the maximum of the pulsed
emission coming from the NS surface remains in phase with the maximum
in the hardest energy band, dominated by the warm spot. 
Therefore, one the two (warmer) regions on the star
surface associated with the magnetic poles must be very close to the
center of the warm spot. It is important to note that, 
while the highest energy bands (dominated by the emission from the single
hot spot), naturally produce produce a single-peaked profile, in the lowest energy band,
dominated by the surface emission (characterized by two symmetrically opposed
warmer regions), a single peak profile is expected only for geometries for which
only one pole is visible as the pulsar rotates (e.g. Page 1995). 

These results and considerations represent the basis to develop a
model that allows us to predict the energy-dependent pulsed fraction
and pulse profile, and use it to determine, for different values of
the NS radius, the viewing geometry and beaming pattern of the emitted
radiation that best match the observations. The model is described
in detail in the next section.

\section{Modeling the spectrum and the pulse profile of \xte\ in quiescence}
\label{emissmod}

Following the motivations given in the previous section, the local temperature
distribution, $T_{\rm th}(r,\theta)$ (in spherical coordinates),
on the surface of \taxp\ is modeled as expected from thermal cooling
under the influence of a dipole magnetic field (Heyl \& Hernquist 1998; 
see also Perna et et al. 2001):
\beq
T_{\rm th}(\theta,\phi)=T_p\left[\frac{4\cos^2\theta_p}{3\cos^2\theta_p+1}\;
(0.75\;\cos^2\theta_p+0.25)^{0.2}\right]\;,
\label{eq:Tdip}
\eeq 
where $T_p$ is the pole temperature and $\theta_p$ is the angle
between the radial direction at position $(\theta,\phi)$ on the surface
of the star and the magnetic pole.

Superimposed to the emission from the whole (cold) surface of the star
is the emission from a small (hot) spot, whose axis coincides with the
dipole axis. 
The mathematical description of the spot on the rotating surface
(of temperature $T_h$ and angular radius $\beta_h$)
follows the theory developed by Pechenick et al. (1993) with some
generalizations presented by Perna \& Gotthelf (2008).
If $\xi$ defines the angle
between the spot/dipole axis and the rotation axis, and $\psi$ the angle
between the observer's direction and the rotation axis, then the angle
$\alpha$ that the axis of the hot spot
makes with respect to the line of sight to the observer can be written as
\beq
\alpha(t)=\arccos(\cos\psi\cos\xi+\sin\psi\sin\xi\cos\gamma(t))\;.
\label{eq:alpha}
\eeq 
This angle is
a function of the phase angle $\gamma(t)=\Omega(t) t$ swept by the star
as it rotates with angular velocity $\Omega(t)$. 
The surface of the star is described by the angular spherical
coordinates $(\theta,\phi)$, and the coordinate system is chosen so
that the $z$ axis coincides with the direction of the line of sight to
the observer. The hot spot is described by the conditions: 
\beq
\theta\le\beta_h,\;\;\;\;\;\;\;\;\;\;\;\; \rmmat{if}\;\;\; \alpha=0\;
\label{eq:con1}
\eeq
and
\beq
   \left\{
  \begin{array}{ll}
    \alpha-\beta_h\le\theta\le\alpha+\beta_h \\
      2\pi-\phi_p^h\le\phi\le\phi_p^h \;\; \;\;\;\;\rmmat{if}
        \;\;\;\alpha\ne 0\;\;\;\rmmat{and} \;\;\;\beta_h\le\alpha\\
  \end{array}\right.\;
\label{eq:con2}
\eeq
where
\beq
\phi_p^h=\arccos\left[\frac{\cos\beta_h-\cos\alpha\cos\theta}{\sin\alpha\sin\theta}\right]\;.
\label{eq:phip}
\eeq

On the other hand, it is identified through the condition 
\beq
\theta\le\theta^h_*(\alpha,\beta_h,\phi),\;\;\;\;\ 
\rmmat{if}\;\;\; \alpha\ne 0\;\;\;\rmmat{and}\;\;\;\beta_h > \alpha\;,
\label{eq:con3}
\eeq
\noindent where the outer boundary $\theta^h_*(\alpha,\beta_h,\phi)$
of the spot is computed by numerical solution of the equation
\beq
\cos\beta_h = \sin\theta_*^h\sin\alpha\cos\phi + \cos\theta_*^h\cos\alpha\;.
\label{eq:t*}
\eeq
Due to the strong NS gravitational field, photons emitted at the NS surface
suffer substantial deflection on their way to the observer.
A photon emitted at a colatitude $\theta$ on the star makes an angle $\delta$ with the
normal to the surface at the point of emission. 
The relation between $\delta$ and $\theta$ is given by the ray-tracing
function\footnote{In the emission code, to improve the computational
efficiency of the above equation, we use the approximation derived by
Beloborodov (2002).} (Page 1995)
\beq
\theta(\delta)=\int_0^{R_s/2R}x\;du\left/\sqrt{\left(1-\frac{R_s}{R}\right)
\left(\frac{R_s}{2R}\right)^2-(1-2u)u^2 x^2}\right.\;,
\label{eq:teta}
\eeq 
\noindent having defined $x\equiv\sin\delta$. Here, $R/R_s$ is the ratio between
the NS and the Schwarzschild radius, $R_s=2GM/c^2$ (we assume $M=1.4
M_\odot$). 

A blackbody model for the local emission is assumed.  While it would
be desirable to perform this analysis with realistic magnetized
atmosphere models, the lack of an extensive set of such models
for high $B$-field strengths, ($B\ge 10^{14}$~G) and arbitrary
  inclinations (with respect to the NS surface) makes this more
complete analysis not yet possible.  This is particularly the case for
the present study, since we are modeling the emission from the entire
surface of the star, and over this there are large regions with a
non-normal $B$.  
 We note, however, recent work extending NS emission models to
  non-normal fields.  In particular, LLoyd (2003a,b) presented model
  spectra for $B \le 10^{14}$~G and for arbitrary orientation, for
  pure Hydrogen composition, and in the limit of complete ionization.
  Ho, Potekhin \& Chabrier (2008) constructed partially ionized
  Hydrogen models for arbitrary field orientation and for strengths in
  the range $10^{12} \le B \le 3\times 10^{13}$~G. While these models are
  very useful for exploring NSs with moderate fields (e.g. Mori \& Ho
  2007), they are still not appropriate for the magnetic field strengths 
  needed for a self-consistent modeling of XTE J1810-197 ($B_{\rm
    dip}$ of about $3 \times 10^{14}$~G).
Hence here we adopt the empirical approach of
parameterizing the level of anisotropy with the function
$f(\delta)\propto\cos^n\delta$ (since 'pencil' beaming dominates in
atmosphere models), and perform the timing analysis with different
values of $n$ within a reasonable range as suggested by beaming in
realistic descriptions of magnetized atmospheres\footnote{In the magnetized
  atmosphere models by van Adelsberg \& Lai (2006), the beaming
  strength depends on the magnitude of the $B$ field, on the
  atmosphere temperature, on the observation energy, as well as on the
  angle $\delta$ itself.  For fields and temperatures in the magnetar
  range, the models by van Adelsberg \& Lai (2006) predict a strong
  forward beaming for angles $\delta\lesssim 40-60$~deg, and a much
  lower anisotropy level (fan-like) at larger angles.  For example,
  for $B\sim 10^{14}$~G and $T\sim 0.4$~keV, an approximation to the
  intensity for $\delta\lesssim 50$~deg is $f(\delta)\propto
  \cos^{0.8}\delta$ at $E\sim 0.2$~keV, and $f(\delta)\propto
  \cos^2\delta$ at $E\sim 1$~keV.  }.

The observed spectrum as a function of phase
angle $\gamma$ is then obtained by integrating the local emission over
the observable surface of the star, including the effect of gravitational
redshift of the radiation (Page 1995)
\begin{eqnarray}
F(E_\infty,\gamma) =\frac{2 \pi}{c\,h^3}\frac{R_\infty^2}{D^2}\;E_\infty^2
e^{-N_{\rm H}\sigma(E_\infty)} \int_0^1 2xdx\nonumber \\ 
 \times \int_0^{2\pi} \frac{d\phi}{2\pi}\; 
I_0(\theta,\phi) \;n[E_\infty e^{-\Lambda_s};T(\theta,\phi)]\;,
\label{eq:flux}
\end{eqnarray} 

\noindent in units of photons cm$^{-2}$ s$^{-1}$ keV$^{-1}$.  In the above
equation, the radius and energy as observed at infinity are given by
$R_\infty= Re^{-\Lambda_s}$, and $E_{\infty}= E e^{\Lambda_s}$, where
$R$ is the star radius, $E$ is the energy emitted at the star surface,
and ${\Lambda_s}$ is defined through the relation  
\beq
e^{\Lambda_s}\equiv\sqrt{1-{\frac{R_s}{R}}}.  
\eeq 

For the spectral function, given by $n(E,T)=1/[\exp(E/kT)-1]$, the
temperature $T(\theta,\phi)$ is equal to $T_h$ if \{$\theta, \phi$\}
satisfy any of the conditions (\ref{eq:con1}) through (\ref{eq:t*}),
and it is given by $T(\theta,\phi)= T_{\rm th}(\theta,\phi)$
otherwise. Correspondingly, the weighted intensity $I_0(\theta,\phi)$
is given by the beaming function $f[\delta(\theta)]$. 

The phase-averaged flux is then readily computed as 
\beq
F_{\rm ave}(E_\infty)=
\frac{1}{2\pi}\int_0^{2\pi}d\gamma F(E_\infty,\gamma)\;.
\label{eq:Fave} 
\eeq
Note that the
 phase dependence $\gamma$ in Eq.(\ref{eq:flux}) comes
from the viewing angles implicit in $\alpha(t)$ and from
the series of conditions (\ref{eq:con1}) through (\ref{eq:t*}).

We also included a multiplicative factor which accounts for the hydrogen column
density between the observer and the star. Note that absorption does influence
the predicted pulsed fractions, when these are computed over finite energy intervals
(Perna et al. 2000).
The magnitude of absorption was fixed to that obtained from the spectral fit (see \S\ref{subs:spec}). 
The model described in this section was imported into the spectral fitting
package XSPEC (Arnaud 1996), and used to fit both spectra and pulse profiles.

\section{Results}
\label{results}

\subsection{Spectral modeling}
\label{subs:spec}

Fitting the phase averaged-spectrum with the model in
Eq.(\ref{eq:Fave}) requires fixing the viewing angles
$\xi,\psi$. However, since the viewing geometry is not known
$a\,priori$, we followed an iterative procedure for our combined
  spectral and timing analysis. We start by assuming $\psi=\xi = 0$,
  the face-on geometry, with the observer looking directly down the
  co-aligned rotation axis and magnetic pole. This implies $\alpha=0$
  and speeds up the spectral fitting substantially as the geometric
  condition describing the hot spot (cfr. Eq.~\ref{eq:con1}) depends
  only on the coordinate $\theta$, reducing the flux integral in
  Eq.(\ref{eq:flux}) to a single dimension. After fitting the
  phase-averaged spectrum, the spectral parameters are used to compute
  the pulse profiles for $\xi,\psi$; this  restricts the range of
  viewing angles $\psi$ and $\xi$ to those that best match the pulsed
  fractions. We then refit the phase-averaged spectrum to refine the
  spectral parameters, but now use the best $\psi,\xi$ from the timing
  analysis. This procedure is iterated until the change in the
  parameters are consistent with the measurement errors. This
  procedure is found to converge in only one iteration and thus
  allows us to explore a wide grid of viewing angles (like e.g. in
  Gotthelf et al. 2010) in a reasonable time.

To begin with, the source distance was first fixed to D=3.3 kpc, based on radio pulse
dispersion measure (Camilo et al. 2006), and consistent with the
measurement derived from HI absorption ($D=3.5\pm0.5$, Minter 2007) 
and the measurement derived from Red Clump Stars in the
direction of the source ($D=3.1\pm0.5$ Durant \& van Kerkwijk 2006).
For completeness, we then also studied the cases corresponding to the
distance fixed at 2 kpc and 5 kpc, which correspond to the borderline
values of the $3\sigma$ uncertainty on the distance.  All the fits
were performed in the 0.5$-$3 keV spectral band. Above 3 keV,
source detection is not significant. 
We repeated the spectral analysis for several values of the beaming
parameter $n$ (in the range $0-2.5$, using steps of 0.5), 
and found that, within the measurement
uncertainties, the inferred spectral parameters were all consistent. Hence,
for the spectral analysis, we set $n=0$. 

Like in the case of the analysis performed by Perna \& Gotthelf
(2008), an important technical issue for these fits is the degree of
degeneracy between the radius $R$ and the variable parameters, in this
case ($kT_h, \beta_h$, $kT_p$).  Without fixing the radius and the
distance there is no unique solution, and we considered a range of
possible values, $9\le R \le 15$~km, in 1~km increments, for the
radius, and the above $3\sigma$ range of $2\le D \le 5$~kpc.

For the distance values of $D=3.3$~kpc and $D=2$~kpc, over the sampled
range of radii, spectral analysis provides acceptable fits, but did
not allow a preferred radius based on the $\chi^2$ measurement (see
table \ref{tab:sp_faceon}).  In the case of $D=5$~kpc, spectral fits
showed a significantly higher $\chi^2$ value, ranging from 1.4 (for
$R=15$~km) to 2.04 (for $R=9$~km). These fits resulted to be
statistically unacceptable; consequently, the timing analysis described in
the following is only performed for $D=3.3$~kpc and $D=2$~kpc.
\begin{table*}
\caption{Spectral fit parameters as a function of the radius R. 
$1\sigma$ c.l. uncertainty is reported.}
\begin{center}
\begin{tabular}{cccccc}
\hline \hline
\multicolumn{6}{c}{$D=3.3$ kpc}\\
\hline
\\        
R  & nH                       & $kT_{\rm BB}$  & $kT_{\rm p}$ & $\beta_{\rm h}$  &   $\chi^2_{\rm red} $   \\ 
km & $\times10^{21}\,cm^{-2}$   & keV          &   keV        & deg          &  (for 75 d.o.f.)         \\
\hline \hline
\\
9 &  $6.9\pm0.1$   &$ 0.57  \pm 0.02  $  & $   0.267 \pm0.001 $  & $2.1 \pm^{0.1}_{0.3}   $   &
1.42 \\

10&  $ 7.1\pm0.1$   &$ 0.54 \pm 0.02   $  &  $  0.250\pm 0.001 $  & $1.8   \pm 0.4  $ &    
1.34 \\

11&  $7.3\pm0.1 $   &$ 0.52 \pm 0.03   $ &   $  0.238 \pm 0.002   $ & $2.0  \pm^{0.6}_{0.1}   $ & 
1.28\\

12&  $7.4\pm0.1 $   &$ 0.47 \pm 0.01  $ &   $  0.226 \pm 0.001 $  & $2.4 \pm^{0.1}_{0.7}$ &  
1.22\\

13&  $7.5\pm0.1 $   &$ 0.45 \pm 0.01 $ &  $ 0.218 \pm 0.001$  & $2.0\pm^{0.3}_{0.5}   $ &   
1.17\\

14&  $7.7\pm0.1 $   &$ 0.44 \pm 0.01 $ &  $0.211\pm0.001 $     & $ 2.1 \pm 0.3 $ &   
1.13\\

15&  $7.8\pm0.1 $   &$ 0.43 \pm 0.01  $ &  $   0.205 \pm 0.001$  & $2.2    \pm^{0.1}_{0.2}   $& 
1.10\\
\hline 
\multicolumn{6}{c}{$D=2.0$ kpc}\\
\hline
9 &  $7.9\pm0.1$   &$ 0.48  \pm 0.01  $  & $   0.235 \pm0.001 $  & $2.6 \pm 0.2   $ &
1.08 \\

10&  $ 8.1\pm0.1$   &$ 0.452 \pm 0.003   $  &  $  0.221\pm 0.001 $  & $2.2   \pm^{0.6}_{0.1} $ &  
1.06 \\

11&  $8.2\pm0.1 $   &$ 0.43 \pm 0.03   $ &   $  0.211 \pm 0.002   $ & $2.5  \pm^{0.2}_{0.1}   $ &  
1.03\\

12&  $8.4\pm0.1 $   &$ 0.416 \pm 0.003  $ &   $  0.202 \pm 0.001 $  & $2.2 \pm 0.2$ &      
1.02\\

13&  $8.5\pm0.1 $   &$ 0.403 \pm 0.002 $ &  $ 0.194 \pm 0.001$  & $2.4 \pm 0.3 $ &   
1.01\\

14&  $8.7\pm0.1 $   &$ 0.391 \pm 0.002 $ &  $0.188\pm0.001 $     & $2.1\pm^{0.6}_{0.1} $ &    
1.00\\

15&  $8.8\pm0.1 $   &$ 0.383 \pm 0.002  $ &  $   0.183 \pm 0.001$  & $2.0    \pm^{0.8}_{0.1}   $&  
1.00\\
\hline \hline         
\end{tabular}
\label{tab:sp_faceon}
\end{center}
\end{table*}

\subsection{Pulsed fraction and pulse profile modeling}
\label{subs:pf}
Given the smooth and nearly sinusoidal pulse shape, 
the pulsed fraction (PF) of
the signal has been determined using the expression
\begin{equation}
{\rm PF}=\frac{F_{\rm max}-F_{\rm min}}{F_{\rm max}+F_{\rm min}}
\label{eq:pf}
\end{equation}
\noindent 
Fluxes are integrated over the given energy bands.
In the adopted geometry, the maximum and minimum fluxes of the model,
$F_{\rm max}$ and $F_{\rm min}$, correspond to phases $\gamma=0$ and
$\gamma=\pi$, respectively. 
As discussed in \S3, this is true for any combination of viewing angles
in the higher energy bands $\ge$ 1keV, in which the emission is either
dominated, or largely influenced,
by the (single) hot spot. In the lowest energy band (0.5-1keV), where
the contribution from the surface emission is dominant, a single
peak can only be obtained for viewing angles $\xi,\psi\lesssim 
\xi_{max},\psi_{max}$. The latter depend mildly on $n,R$, and are
generally $\lesssim 50^\circ-55^\circ$. Our viewing parameter search is restricted
to the range of angles for which the lowest energy band remains single peaked.
Incidentally, the high $\xi,\psi$ parameter range is also {\em independently} 
ruled out by the PF in the highest energy band alone: for most combinations of $n,R$, 
the predicted PF would be much higher than the measured value. 

A measure of the PF in each energy interval was obtained
using the timing solution, phase-connecting
the three pointings.  The background level, which was variable during
the three pointings, was subtracted for each
multi-instrument lightcurve.
We found that the PF
increases with energy: $\rm PF_{(0.5-1\, keV)}=17\pm1\%$,
$\rm PF_{(1-1.5\, keV)}=26\pm0.8\%$, $\rm PF_{(1.5-2\,
keV)}=36\pm1.3\%$, $\rm PF_{(2-3\, keV)}=47\pm2.6\%$ (all the
uncertainties are hereafter reported at $1\sigma$ confidence level,
c.l., unless otherwise stated). Above 3~keV, due the low S/N ratio,
only an upper limit on the PF could be obtained (but this is not an
useful value for the purpose of this work). Hence our timing analysis
uses the data up to 3~keV. 

Starting with the best spectral fit model parameters presented in
Table \ref{tab:sp_faceon}, obtained for $\psi=\xi=0$, the best values
of $\psi^{*}$ and $\xi^{*}$ needed to reproduce the observed pulsed fraction 
were searched (in the case of D=3.3 kpc and
D=2 kpc).  Given the high S/N of the data, both the
pulsed fraction and the full pulse profile were modeled.

For each value of the NS radius fitted for in \S\ref{subs:spec} (9--15
km), the pulsed fraction, defined in Eq.(\ref{eq:pf}), was computed
over the grid of angles $(\xi,\psi)\le (\xi_{\rm max},\psi_{\rm max})$
deg, in 1 degree intervals.  For each value on the grid, the model
predictions were compared with data. Notice that the flux depends on
the angles $\xi$ and $\psi$ only through the parameter $\alpha$ in
Eq.(\ref{eq:alpha}), and therefore it is symmetric with respect to an
exchange of $\xi$ and $\psi$.

In order to explore the behaviour of the PF with both $R$ and $n$ independently,
we first examined how the PF varied with $R$ at fixed $n$, and then
how they varied with $n$ at fixed $R$. In the following, we first report the
results for $n=1$ and $R=9-15$~km, in steps of 1~km and then for
$R=9$~km and $R=15$~km, for $n=0-2.5$, in steps of 0.5. 
The behaviour for other combinations of parameters
can then be inferred from the results shown.

\begin{figure*}
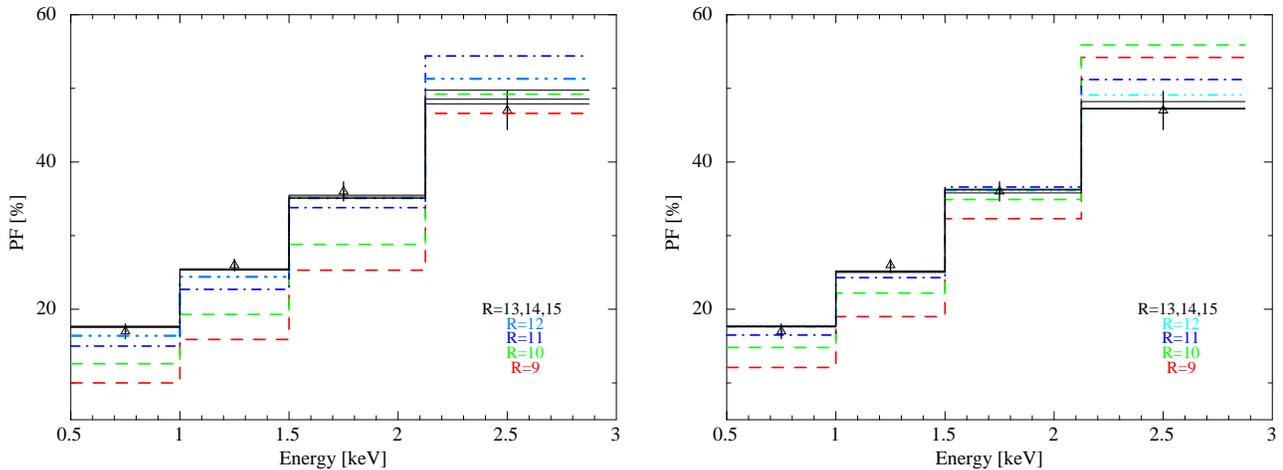

\begin{center}
\begin{tabular}{cc}
\includegraphics[angle=-90,scale=.34]{figpftalk.ps} & \includegraphics[angle=-90,scale=.34]{pfmodvsdata_2kpc.ps}\\
\end{tabular}
\caption{Model predicted PF (for the best emission geometry
  as reported in Table \ref{tabpf}) for different radius values,
  compared with data. Left panel is for $D=3.3$ kpc, while right panel
  is for $D=2.0$ kpc. Beaming factor is equal to 1. Dashed and
  dot-dashed lines represent the model prediction for low radius
  values ($9\leq R \leq12$ km), while continuous black lines are
  obtained for high radius values ($13\leq R \leq15$ km). $1\sigma$
  c.l. uncertainty is reported.}
\label{fig:pf_modvsdata}
\end{center}
\end{figure*}

For the case $n=1$, 
Table \ref{tabpf} reports the best geometry
(expressed through the angles $\psi^{*}$ and $\xi^{*}$), for each sampled
radius, and the corresponding predicted PF as a function of the energy
band, in comparison with the data. For this case, we further explored
the dependence of the results on the source distance. 
In the case of $D=3.3$ kpc, low radius values ($9\leq R \leq12$ km),
underpredict the observed PF for low energy intervals $0.5\leq E
\leq2$ keV. Moreover, for $R=11$ and $R=12$ km, the model overpredicts the
observed PF in the highest energy band ($2\leq E \leq3$ keV; see \ref{fig:pf_modvsdata}).
The different trend of the PFs with radius in the low and high energy
bands is due to the fact that the flux in these is dominated by
different components: the cooler component from the all NS surface
in the low energy band, and the hotter small spot in the high energy band.
The smooth temperature gradient of the former makes the PFs less sensitive
to changes in the viewing geometry, and hence the dominant factor in determining
the change in the PF with $R$ is the general-relativistic suppression of the
pulsed fraction as $R$ decreases. On the other hand, the pulsed fraction produced 
by the small hot spot is much more sensitive to changes in viewing angles
with the result that smaller $R's$, which are best fit 
by larger values of $\xi,\psi$ (see Table \ref{tabpf}), yield higher PFs. 
 We found that it is not possible to reproduce the
observed pulsed fraction for low radius values ($9\leq R \leq12$ km),
which are hence rejected by the model for this distance. The predicted
PF for high radius values ($13\leq R \leq15$ km) is instead fully
consistent with the data, within $1\sigma$
uncertainty. Figure~\ref{fig:pf_modvsdata} shows the model predicted
PF, for all the values of the radii considered, compared with the
data.  Similar results are obtained for $D=2$ kpc; however in this case
the range of allowed radii is wider, including also $R=12$ km.
For, e.g., an intermediate value of the radius, $R=13$ km ($D=3.3$ kpc),
the best viewing geometry angles range from
$\psi^{*}=\xi^{*}=38^{\circ}$ to
($\psi\,,\xi)=(52^{\circ}\,\,29^{\circ}$) at $3\sigma$ c.l. (see
Table \ref{tabpf}).   The angles ($\psi\,,\xi$) that provide the best
match to the PF data clearly vary with $R$, for a fixed value 
of the beaming factor $n$. Smaller radii require a
larger variation in $\alpha(t)$ ($\alpha$ varies between 
$\psi-\xi$ and $\psi+\xi$) to compensate for the stronger
general-relativistic suppression of the flux modulation.

\begin{figure}
\begin{center}
\includegraphics[width=3.4in, height=3.4in, angle=0]{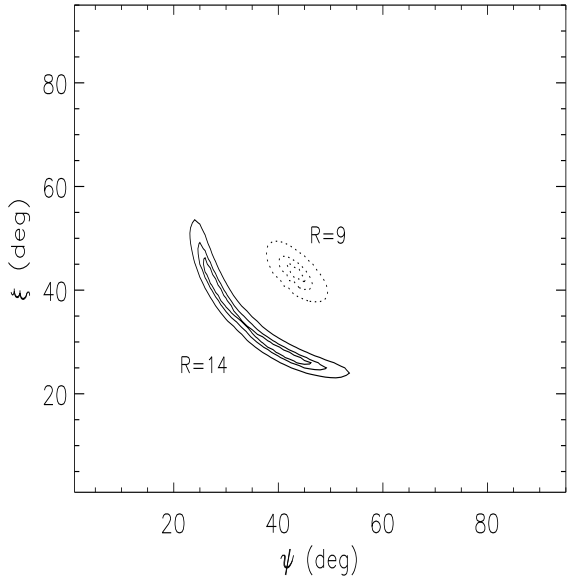}
\caption{Reduced chi-square ($\chi_{\nu}^2$) maps obtained by
comparing data and model pulsed fraction described in the text for a range
of viewing angles $\psi$ and $\xi$. The 68\%, 90\% and 99\% confidence
levels are shown for the best match to the observed pulsed fractions
using the beaming pattern $n=1$ for $R=9$~km and $R=14$~km. 
The results are clearly degenerate with respect to an interchange of
$(\xi,\psi)$. A comparison with Fig.~2 in Perna \& Gotthelf (2008)
shows the substantial reduction in the confidence range.
} 
\label{fig:contour}
\end{center}
\end{figure}

The $\chi^2_{\nu}$ map, computed for both the case of $R=9$~km 
and $R=14$~km (Figure \ref{fig:contour}), displays the 68\%, 90\% and 99\% confidence
levels. This map is produced by comparing the
model and observed pulsed fraction over a range of possible ($\xi,\psi$)
angle pairs, for our best fit spectral model parameters. The viewing
geometry was found to be well constrained, with the range of allowed
$3\sigma$ solutions a factor of 2.5 smaller than (but consistent with)
what was found by Perna \& Gotthelf (2008) in the analysis of the outburst decay  
(where the $3\sigma$ c.l. of the viewing parameters were
ranging from $\psi^{*}=\xi^{*}=37^{\circ}$ to $\psi,\xi=85^{\circ},15^{\circ}$).
This consistency is not surprising, since the warm spot appears to be the remnant
of the heated regions during the outburst. 
The elongated shape of the contour plots shows that the two angles $\psi$
and $\xi$ are highly correlated in the fit. This is a result of the
fact that the PF depends on a combination of these two angles.

\begin{figure}
\begin{center}
\includegraphics[angle=-90,scale=.35]{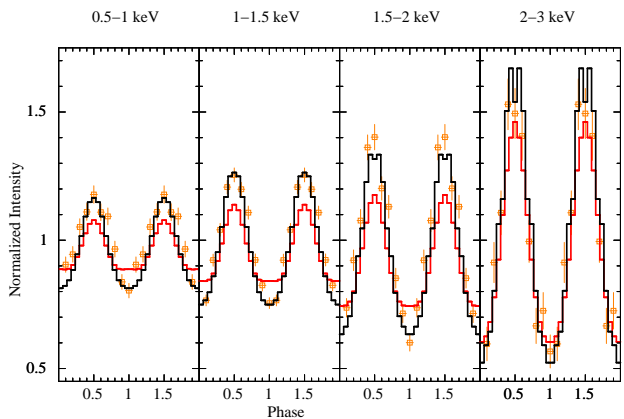}
\caption{Predicted pulse profile for $R = 13$~km (black line), and $R = 9$~km
(red line), $D=3.3$ kpc and $n=1$,  in four energy bands, compared to
  data (orange squares).  The two cases $R=14$ and $R=15$~km give fully
  consistent results with $R = 13$~km.
Each profile is computed with the best fit viewing angles as reported in Table 2. 
}
\label{fig:confshape}
\end{center}
\end{figure}

\begin{table*}
\caption{Best emission geometry angles $\psi^{*}$ and $\xi^{*}$ and corresponding predicted
pulsed fraction, for $n=1$ and variable $R$, compared to the data.
$1\sigma$ c.l. uncertainty is reported}
\begin{center}
\begin{tabular}{ccccccc} 
\hline \hline
\multicolumn{6}{c}{$D=3.3$ kpc}\\
\hline
\\
R&$\rm PF_{0.5\div1\, keV}$&$\rm PF_{1\div1.5\, keV}$&$\rm PF_{1.5\div2\, keV}$&$\rm PF_{2\div3\, keV}$& $\psi^{*}$,$\xi^{*}$\\
km    &\% &\% &\% &\% &  degree    \\
\hline
9&   10.0   & 15.7      & 24.9  &  45.8  & 48,48  
\\
10& 12.6   &  19.3   &  28.8  & 49.2  & 46,46  
\\
11& 15.0 &  22.7     & 33.8  &   54.4 & 44,44  
\\
12& 16.6  &  24.7    & 35.6  &  52.1 & 40,39  
\\
13& 17.5  &  25.3    & 34.9 &  49.8   & 38,38  
\\
14& 17.5    &  25.2    & 34.8 &  48.1 & 40,31  
\\
15& 17.7    &  25.4    & 35.3 &  47.8 & 41,28 
\\
\hline
\multicolumn{6}{c}{$D=2.0$ kpc}\\
\hline
9& 12.1    &  19.0     &  32.3 & 54.2 & 46,46 
\\
10& 14.8   &  22.2   &  34.9  & 55.9 & 44,44 
\\
11& 16.5  &  24.3     & 36.6  &  51.2  & 39,38  
\\
12& 17.6   &  25.1   & 36.1  &  49.1 & 37,36  
\\
13& 17.6 &  25.2    & 36.3  &  47.2   & 41,28  
\\
14& 17.7   &  25.1    & 36.2 &  48.2 & 45,25 
\\
15& 17.7    &  25.0    & 35.8 &  47.3 & 46,23  
\\
\hline
observed PF &  $17   \pm 1.0  $         &  $ 26.0\pm0.8 $             &   $36\pm1.3$        &   $47\pm2.6    $        &  
\\
\hline \hline
\end{tabular}
\label{tabpf}
\end{center}
\end{table*}

 Having assumed a face-on spectrum to derive the spectral model
  parameters used to compute the pulsed fractions, we now proceeded to
  refit the spectrum, for each radius, using the best viewing geometry
  $\psi^*, \xi^*$.  This is iterated to a convergence criteria set by
  the measurement errors.  The final spectral values in all cases are
  consistent with those found for $\psi=\xi=0$. (Table 1).  For
  example, for the special cases of $R=13$~km, $D=3.3$~kpc, and
  $\psi^{*}=38^{\circ}$ and $\xi^{*}=38^{\circ}$, we find $n_{\rm
    H}=7.5\pm 0.1$, $kT_{\rm BB}=0.453\pm0.003$, $kT_{\rm p}=
  0.218\pm0.001$ and $\beta_h=2.0\pm^{0.2}_{0.1}$, with $\chi^2=87.7$
  for 75 d.o.f. (see Figure \ref{fig:spec13}); as another example, for
  $R = 13$~km, $D = 2.0$~kpc, and $\psi^{*}=42^{\circ}$ and
  $\xi^{*}=28^{\circ}$, we find $n_{\rm H}=8.4\pm 0.1$, $kT_{\rm
    BB}=0.413\pm 0.003$, $kT_{\rm p}= 0.24\pm 0.03$ and
  $\beta_h=2.0\pm^{0.2}_{0.1}$, with $\chi^2=76.4$ (for 75 d.o.f.).
Given the consistency
  between the spectral parameters determined with $\psi=\xi=0$ and with
  those for the best viewing angles $\psi^*,\xi^*$, the iterative process
  did not need to be continued further.  For the
  best fit spectral parameters and viewing angles, the resulting
  pulsed fractions and pulse profile, for several radii, are show in
  Figure 3 and 5, respectively. It is important to note that, for a
  pre-determined emission pattern of the radiation (here parameterized
  by the beaming parameter n), there are values of radii (e.g., $R\le 12$~km
  for $n = 1$) for which no good solution to the combined
  spectral/timing properties of the source can be found. Hence this
  type of analysis provides useful limits on the NS radius in the
  context of the model.

\begin{figure*}
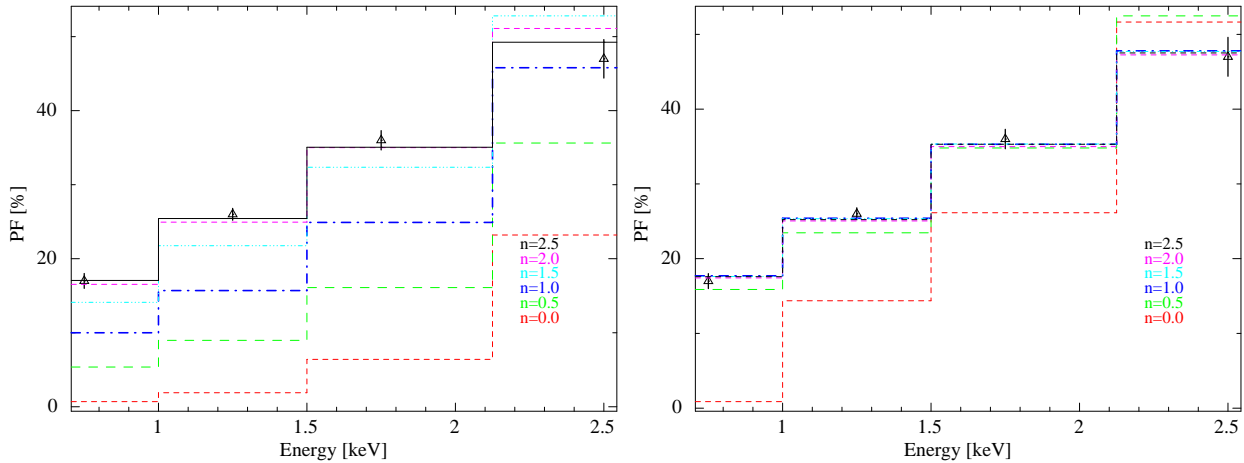

\begin{center}
\begin{tabular}{c}
\includegraphics[angle=-90,scale=.34]{pfvse_n.ps} 
\includegraphics[angle=-90,scale=.34]{pfR15vse_n.ps}\\
\end{tabular}
\caption{Model predicted PF (starting from the best emission geometry as reported in Table \ref{tab:pfvsn}) 
for different values of the beaming pattern $n$. Left panel is for R=9 km, while right panel is for 
R=15 km. $1\sigma$ c.l. uncertainty is reported.} 
\label{fig:pf_n}
\end{center}
\end{figure*}

\begin{figure*}
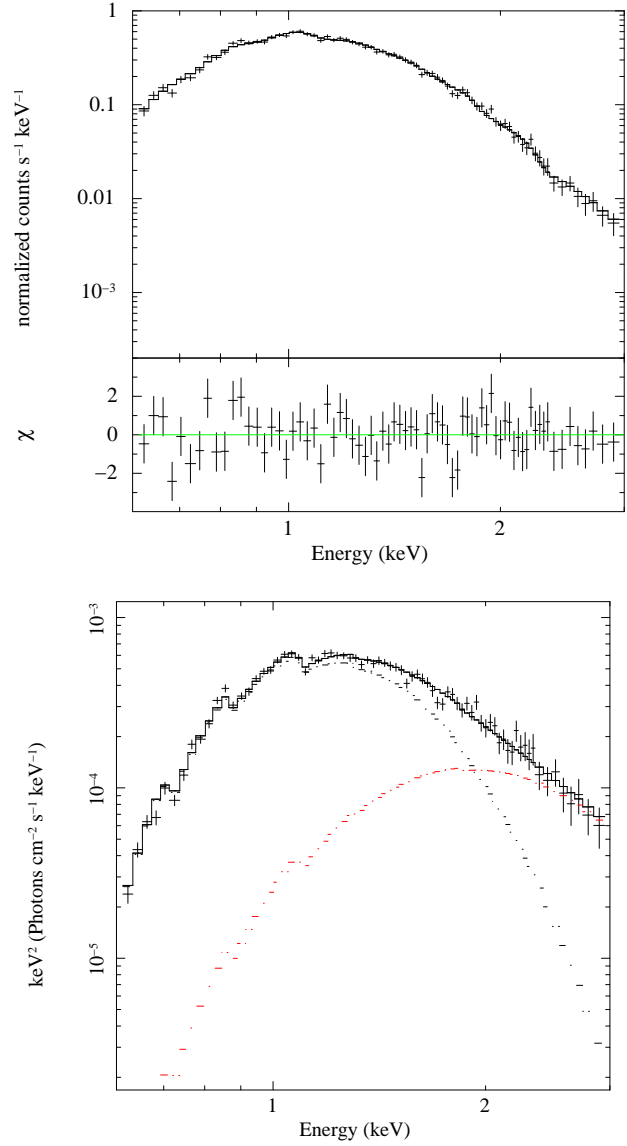

\begin{center}
\begin{tabular}{cc}
\includegraphics[angle=-90,scale=.43]{d2.ps}  \\
\includegraphics[angle=-90,scale=.41]{fig7a.ps} \\
\end{tabular}
\caption{\textit{Upper panel}: \XMM\ phase-averaged spectrum of \xte\ obtained with the model
presented in the text, for the specific case
of R=13 km, D=3.3 kpc, and the best fit angles $\psi^{*}=38^{\circ}$ and $\xi^{*}=38^{\circ}$. 
Model residuals are shown in the lower panel.
The parameters for the best fit model are reported in Table
\ref{tab:sp_faceon}. The data show the summed spectra from the three observations.
\textit{Lower panel}: Same as the upper panel, but for unfolded source spectra. The contribution
of the different
model component is also shown: surface (black dotted line), warm spot (red dotted line).}
\label{fig:spec13}
\end{center}
\end{figure*}

To consider the effect of the angular distribution of radiation on our spectral and timing
modeling, we repeated the above analysis for various assumed beaming indexes. Table~\ref{tab:pfvsn}
reports the model predicted PF for $0\leq n\leq2.5$, in 0.5 step increments, for $R=9$~km
and $R=15$~km (see Figure~\ref{fig:pf_n}). For $n=0$ (isotropic
emission), we found that not even the largest radii were able to
account for the observed level of modulation.  The modeled PF is, in
fact, produced by the interplay of two flux components: the one from
the whole NS surface, and the other from the small, hot spot. The
high level of pulsation which could have been produced by
the hot component alone, is strongly reduced by the presence of
emission on the entire NS surface.  If the local emission is
isotropic, not even the largest radii can provide sufficiently large
PFs. However, as the value of $n$ increases, the range of allowed
radii becomes wider; we found that for $n=0.5$, only $R=15$~km
provides a marginal match for the energy-dependent PF; for $n=1$, the
range $R\geq 13$~km provides a good overall representation of the
data.  For $n=2.5$, due to the resulting strong increase of the PF at
all energies, a good match to the PFs can be found for each value of
$R$ (but clearly for different combinations of the angles
$\psi,\xi$). While our results provide a hint to the presence of a
magnetized atmosphere on the NS surface, they also show that without a
priori knowledge of $f(\delta)$ for each ($\theta,\phi$) on the NS
surface, the timing analysis does not allow to break the degeneracy
with the radius (at least within our measurement uncertainties).

\begin{table*}
\caption{Best emission geometry angles $\psi^{*}$ and $\xi^{*}$ and corresponding predicted
pulsed fraction, for different value of the beaming factor $n$ and $D=3.3$ kpc, compared to data. 
$1\sigma$ c.l. uncertainty is reported}
\begin{center}
\begin{tabular}{cccccc} 
\hline \hline
\multicolumn{6}{c}{$R=9$ km}\\
\hline
\\
$n$&$\rm PF_{0.5\div1\, keV}$&$\rm PF_{1\div1.5\, keV}$&$\rm PF_{1.5\div2\, keV}$&$\rm PF_{2\div3\, keV}$& $\psi^{*}$,$\xi^{*}$\\
    &\% &\% &\% &\% &  degree  \\
\hline
0.0&   0.7   & 2.0      & 6.4  &  23.2  & 55,55    
\\
0.5& 5.3   &  9.0   &  16.1  & 35.6  & 50,50  
\\
1.0& 10.0 &  15.7     & 24.9  &   45.8 & 48,48  
\\
1.5& 14.1  &  21.8    & 32.4  &  52.8 & 45,45   
\\
2.0& 16.5  &  24.9    & 35.0 &  51.1   & 39,38  
\\
2.5& 17.1    &  25.4    & 35.1 &  49.2 & 37,31    
\\
\hline
\multicolumn{6}{c}{$R=15$ km}\\
\hline
0.0& 8.8    &  14.4  & 26.1  & 51.7 &  51,48 
\\
0.5& 15.9   &  23.5   & 34.8  & 52.5 & 42,41  
\\
1.0& 17.7  &  25.4    &  35.3 & 47.8  & 41,28  
\\
1.5& 17.6   &  25.3   &  35.3 & 47.7 & 44,21   
\\
2.0& 17.4  &   25.0    & 35.0   & 47.2 & 43,18  
\\
2.5& 17.6   & 25.2    & 35.3  & 47.5 &  38.18  
\\
\hline
observed PF &  $17   \pm 1.0  $         &  $ 26.0\pm0.8 $             &   $36\pm1.3$        &   $47\pm2.6    $        &  
\\
\hline \hline
\end{tabular}
\label{tab:pfvsn}
\end{center}
\end{table*}

\section{Summary and discussion}

The decline of the post-outburst emission of \xte\  has
allowed us to perform a detailed spectral and timing analysis of the
emission of this source in its quiescent state.  
The main results from the modeling described herein can be summarized as follows: 

\begin{enumerate}

 \item The temperature distribution on the surface of the NS is
   consistent with the expectations of a dipole magnetic field
   configuration.  We remark that our analysis does not
   guarantee that a dipolar magnetic field  represents the
   only possible solution; in principle, other magnetic field
   configurations and beaming patterns can be ad-hoc produced and
   tested for this source. However, a dipolar field is the
   simplest magnetic field configuration for a pulsar, and one that
   yields a single peak, nearly sinusoidal (as observed in \xte), for a
   wide range of viewing angles. 
   Hence, although not formally unique from a mathematical
   point of view, the proposed solution is physically motivated, and
   the fact that it provides such a good match to the data yields
   confidence that it is indeed a reasonably good representation of the
   quiescent emission of this transient magnetar.  
\item The pulsed fraction of the NS emission requires an anisotropic,
  pencil-type radiation pattern, which is a likely indication of the
  presence of a magnetized atmosphere on the NS surface.  For a {
    $\cos^{n}\delta$} emission profile, we performed a timing analysis
  for $n$ in the range 0-2.5 (in steps of 0.5), and radii between
  9-15~km (in steps of 1~km).  We found that no match to the PF data
  could be obtained for $n=0$ (isotropic radiation pattern), no matter
  the value of the radius, while for $n=2.5$, a good match to the PFs
  could be found even for $R=9$~km (smaller values of the radius are
  increasingly allowed as $n$ increases).  For each value of $n$ in
  between there is a range of radii which are not allowed by the PF
  data (e.g. for $n=1$ no good match could be found for $R\le 12$~km).
Therefore, our analysis has clearly demonstrated how a detailed {\em a
    priori} knowledge of the spectral distribution and emission
  pattern as a function of ($\theta, \phi$) on the entire surface of
  the star has the power to allow a radius constraint for the NS.

 \item The overall emission geometry is constrained by identifying
   likelihood regions in the $\psi-\xi$ parameter space. The
   most significant range according to our fits is
   consistent with the results of the earlier, post-outburst analysis
   of Perna \& Gotthelf (2008), but the $3\sigma$σ allowed region is now
   reduced by a factor of 2.5. We note that the hot-spot axis used in
   the earlier work coincides with the dipole axis in the current
   study.

A spectral and timing analysis of the post-outburst emission of \xte\,
has been performed also by Albano et al. (2010). They fitted the first
7 observations using a three-temperature model. The X-ray emission is
produced in a globally twisted magnetosphere, and parameterized by the
twist angle, the electron velocity and the seed photon temperature. In
the the last two sets, the flux is approaching the quiescent level,
and only two temperatures are required to fit the X-ray spectrum.  For
these observations, they found that the angle between the magnetic
axis and the rotation axis is $\xi= {30.0^\circ}^{+12.3}_{-20.0}$ in
the 6th observation, and $\xi= {22.7^\circ}^{+16.4}_{-20.0}$ in the
7th, 6 months later, while the angle between the line of sight and the
rotation axis was found to be $\psi= {153.9^\circ}^{+19.6}_{-16.0}$ in
the 6th and $\psi= {145.8^\circ}^{+16.4}_{-9.5}$ in the 7th
observation.  While the values of the angle $\xi$ are roughly within
the range of what we find, the angle $\psi$ is inconsistent with
ours. A direct comparison between our modeling and theirs is however
not possible due to some fundamental differences in the basic
assumptions. Our quiescent emission is assumed to be thermal, and we
leave the anisotropy level of the local radiation as a free
parameter. On the other hand, the twisted magnetosphere model has a
well determined angular radiation pattern. The NS surface temperature
is assumed to follow a dipolar-like pattern in our model, while in
theirs it is assumed to be constant.  Furthermore, our modeling
includes the general relativistic effect of light deflection (which
heavily influences the pulsed fractions and hence the viewing geometry
that we infer), while their modeling does not.

\end{enumerate}

Our study has allowed us to establish some important properties
of the quiescent emission of \xte\,. The spectral analysis has
shown evidence for the presence of a very small hot spot, only
a couple of degrees in size. This emission component dominates the X-ray 
flux above about 1 keV. Interestingly, a small hot region is also
characteristic of the quiescent emission of another (transient) AXP,
\cxou\, (Israel et al. 2007; Skinner et al. 2006). The most natural
interpretation, in the case of \xte\, as well as in the case of \cxou\,, is
the association of this warmer region to a concentration of magnetic field
lines, since the surface heat flux of a magnetar scales
sensitively as $B^{4.4}$ (Thompson \& Duncan 1996). It is then not surprising
that this region coincides with the location of the outburst, as our analysis
for \xte\, has shown. 

Our constraints on the viewing geometry could in principle be combined
with those obtained from radio measurements following the outburst, to
further restrict the allowed parameter space. Camilo et al. (2007a)
showed that the radio and X-ray locations were aligned, within some
rather large uncertainty due to the small number of counts of their
{\em Chandra} data. Camilo et al. (2007b) used radio polarimetry to
fit for the angle angle between the magnetic field and the rotation
axis ($\xi$ in our notation), and for the angle between the magnetic
field axis and the line of sight ($\alpha_{min}=\psi-\xi$ in our
notation), finding two possible configurations: one with
$\xi=70^\circ$ and $\alpha_{min}\sim 20^\circ-25^\circ$, and another
with $\xi=4^\circ$ and $\alpha_{min}=4^\circ$.  Polarization
observations of \xte\, were also used by Kramer et al. (2007) to
constrain the viewing geometry of this pulsar.  They identified a main
pulse and a mid pulse, produced in different locations on the star.
Their best fitted angles (with the same notation as above) were
$\xi=44^\circ$ and $\alpha_{min}=39^\circ$ for the main pulse, and
$\xi=76^\circ$ and $\alpha_{min}=6^\circ$ for the inter pulse.  Given
some discrepancies between the different angle estimates from radio
data, and the lack of a firm and accurate association between the
radio and X-ray emitting regions, we refrain in the current work from
using the radio angle determinations to restrict our allowed
parameters range. However, we take the opportunity to emphasize the
importance of simultaneous radio-X ray observations should another
outburst occur (either in this or in another magnetar).

Our analysis has allowed us to explore the consistency of the data
with the expectations for the temperature distribution on the NS
surface if the $B$ field is dominated by a dipolar component.  The
angles $\xi$ and $\psi$ that provide a good match to the observed
pulsed fractions (Fig.~3 and Table~2) also ensure that the profile
remains single-peaked in the lowest energy band, dominated by the
surface/dipolar component. Combinations of the beaming parameter and
the NS radius can be found for which our model provides an excellent
match to spectra, pulsed fractions and, more generally, to the full
pulse profile.  Within the context of magnetars, being able to infer
the magnetic topology is important for a number of reasons. Firstly,
measurements of the magnetic field strength from $P$ and $\dot{P}$
make the implicit assumption of a dipolar field; substantial
departures from this configuration will result in biased estimates.
Second, the magnetic field topology in magnetars also plays an
important role in the resulting outbursting behaviour (Thompson \&
Duncan 1995, 1996), in that it regulates the intensity, frequency and
location on the NS surface of the starquakes, and hence of the
outbursts (Perna \& Pons 2011).

Ultimately, we note how the combined spectral/timing analysis that we
performed has the potential to constrain the compactness ratio of the
neutron star (or equivalently the radius, for a given mass). However,
obtaining this type of constraint requires an a priori knowledge of
the spectral and radiation pattern of the local emission.  This is
particularly difficult in the case of highly magnetized objects, since
the local emission depends on the local strength and orientation of
the $B$ field, which in turn would need to be determined as a part of
the fitting procedure. This analysis,  which has begun for NSs
  with magnetic fields in the $10^{12}-10^{13}$~G range (Ho \&
  Mori 2008), will become possible for magnetars once atmospheric
models for arbitrary field orientations  (Lloyd 2003a,b; Ho et
  al. 2007) are extended to strengths $B\sim 10^{14}-10^{15}$~G.

\section*{Acknowledgements}

Federico Bernardini thanks Prof. Svetozar Zhekov for his precious help. This work was
was partially suppported by grants NSF-AST 1009396, NASA NNX10AK78G, NNX09AT17G, NNX09AT22G,
NNX09AU34G, GO0-11077X, DD1-12052X, G09-0156X, AR1-12003X, DD1-12053X (RP).

\label{lastpage}
\end{document}